# The density and disorder tuned superconductor-metal transition in two dimensions

Zhuoyu Chen[1,2,3], Adrian G. Swartz[1,2,3], Hyeok Yoon[1,2], Hisashi Inoue[1,2,3], Tyler Merz[1,2], Di Lu[1,3,4], Yanwu Xie[1,2,3], Hongtao Yuan[1,3], Yasuyuki Hikita[1,3], Srinivas Raghu[1,3,4], Harold Y. Hwang[1,2,3]

[1]*Geballe Laboratory for Advanced Materials, Stanford University, Stanford, CA 94305, USA.*

[2]*Department of Applied Physics, Stanford University, Stanford, CA 94305, USA.*

[3]*Stanford Institute for Materials and Energy Sciences, SLAC National Accelerator Laboratory, Menlo Park, CA 94025, USA.*

[4]*Department of Physics, Stanford University, Stanford, CA 94305, USA.*

**Quantum ground states which arise at atomically controlled oxide interfaces provide an opportunity to address key questions in condensed matter physics, including the nature of two-dimensional (2D) metallic behaviour often observed adjacent to superconductivity**[1–8]**. At the superconducting LaAlO$_3$/SrTiO$_3$ interface**[9,10]**, a metallic ground state emerges upon the collapse of superconductivity with field-effect gating**[11,12]**. Strikingly, such metallicity is accompanied with a pseudogap**[13]**. Here, we utilize independent control of carrier density and disorder of the interfacial superconductor using dual electrostatic gates**[14]**, which enables the comprehensive examination of the electronic phase diagram approaching zero temperature. We find that the pseudogap corresponds to precursor pairing, and the onset of long-range phase coherence forms a 2D superconducting dome as a function of the dual gate voltages. The gate-tuned superconductor-metal transitions are driven by macroscopic phase fluctuations of Josephson coupled superconducting puddles.**

General scaling arguments indicate that metallic states are absent at zero temperature in weakly interacting and weakly disordered 2D electron systems[15]. Nevertheless, apparent metallic ground states in proximity to 2D superconductivity have been observed in various experiments[1–8]. Thus, the existence and nature of the 2D quantum (zero-temperature) metal has been a matter of long-standing debate[16–21]. At the electrostatically-gated LaAlO$_3$/SrTiO$_3$ interface 2D superconductor, a superconducting dome is formed with the appearance of a pseudogap on the "underdoped" side[11–13]. Interestingly, metallic behavior emerges upon quenching of the 2D superconductor, an aspect which has generally not been emphasized in previous reports on LaAlO$_3$/SrTiO$_3$. The coexistence of metallic and pseudogap behaviour suggests that further examination of the connection between the superconducting and metallic ground states would be insightful. This calls for experiments that can independently manipulate the key parameters of disorder and carrier density. Previous experiments, using a single gate from the back of the undoped-SrTiO$_3$ substrate, simultaneously influences both the carrier density and interface scattering from the tail of the electron distribution envelope[12,14,22], making a disentanglement of singly-gated behavior difficult. Here, we implement dual electrostatic gates, providing systematic control over the effective interfacial disorder, carrier distribution, and density, enabling a wide-range mapping of the low-temperature phase diagram.

First, we describe the fundamental operation of the dual electrostatic gates[14] (structure shown in Fig. 1**a**), which are used to manipulate the electrostatic boundary conditions of the asymmetric quantum well within SrTiO$_3$ (Fig. 1**b**,**c**). The confinement potential is bounded sharply at the interface by the wide-gap insulator LaAlO$_3$ but extends gradually into the quantum paraelectric SrTiO$_3$ following an electric field dependent dielectric function. Carriers close to the interface experience higher scattering, compared to carriers extending further into the clean SrTiO$_3$ substrate. Voltages applied to the top-gate ($V_{TG}$) predominantly control the density of carriers confined close to the interface, but the electronic thickness remains nearly constant (Fig. 1**b**). When $V_{TG}$ is decreased, the overall scattering is reduced due to the decrease of carriers distributed close to the interface, giving rise to an increase of mobility for $V_{TG} > 0.6$ V in Fig. 1**d**. (Below 0.6 V in Fig. 1**d**, localization behaviour appears at low carrier density, along with reduced mobility.) In contrast, the back-gate primarily controls the tail of the quantum well extending into the SrTiO$_3$ substrate. With decreasing back-gate voltage ($V_{BG}$), the thickness of the conduction layer decreases dramatically such that the narrowly confined carriers are strongly scattered by the interfacial disorder (Fig. 1**c**) and the mobility decreases (Fig. 1**e**). Importantly, the large, and strongly electric field dependent dielectric response of SrTiO$_3$ greatly enhances the dual-gate tunability compared to conventional semiconductor heterostructures[23]. A complete presentation of dual-gating normal state transport, and Poisson-Schrödinger simulations in the presence of the nonlinear dielectric response, can be found in ref. 14.

Next, we discuss the top-gate modulation of the interface ground states. Figure 2**a** plots the resistivity versus temperature (*R-T*) curves with varying $V_{TG}$ and fixed $V_{BG} = 0$ V. In the low $V_{TG}$ regime, the resistivity upturns slightly as temperature decreases, which may indicate an initial tendency toward localization. With increasing $V_{TG}$, the resistivity at $T = 400$ mK (denoted as $R_N$) decreases and crosses $h/e^2 = 26$ kΩ/sq. In this regime, the *R-T* curves are nearly temperature independent. We note that this behaviour is rather different from metal-insulator transitions in conventional semiconductor systems[20,21]. When $R_N$ falls below ~ 2 kΩ/sq., the resistivity exhibits a drop with decreasing temperature, followed by a saturating finite resistance approaching zero temperature (also Fig. 2**a** inset). The saturating resistivity at the lowest measured temperature can be as much as two orders of magnitude lower than $R_N$. Upon further increasing $V_{TG}$, the resistivity drops below the measurement noise limit, indicating macroscopic superconductivity. We take here a functional definition of the superconducting transition temperature ($T_C$) as the temperature at which the resistivity drops below 1% of $R_N$. Importantly, we find that the *R-T* curves exhibit a two-step feature upon transitioning into the superconducting state. These sequential resistive drops with decreasing temperature can be identified by peaks in the second derivative of the *R-T* curves (denoted as $T_P$ and $T_F$ in Fig. 2**b**). The top-gate tuned phase diagram with four distinct regimes is thus obtained by taking $T_P$, $T_F$, and $T_C$ as boundaries as shown in Fig. 2**c**. We will discuss the assignment of these states in detail below, but note here that the boundaries of the phase diagram extrapolate to zero temperature, implying different ground states.

The counterpart $R$-$T$ curves and phase diagram under varying $V_{BG}$ at fixed $V_{TG}$ = 1.8 V is shown in Fig. 3. Following the analysis of the second derivative of the $R$-$T$ curves (Fig. 3**a** and 3**b**), a phase diagram is obtained as a function of $V_{BG}$ (Fig. 3**c**). Note that back-gate modulation produces a rather different phase diagram from the top-gate case (shown previously in Fig. 2): $T_P$ monotonically decreases with $V_{BG}$; both $T_F$ and $T_C$ are non-monotonic, and $T_C$ exhibits a complete dome; on both sides of the dome, the resistivity saturates at finite values approaching zero temperature[12]. A key finding here is the deviation between $T_P$ and $T_C$ on the "underdoped" (negative back-gate) side of the dome. The behavior of $T_P$ and $T_C$ is qualitatively consistent across the dome with the pseudogap observations from tunneling spectroscopy[13]. This implies a correspondence between $T_P$ and gap opening.

These results demonstrate that a variety of emergent ground states are achievable solely by electrostatic gates. To understand their nature, we first discuss the different length scales for disorder, inhomogeneity, and superconductivity at the interface. Due to finite interdiffusion, the interface where the 2D superconductor resides has disorder on a range comparable to the lattice constant[13,24]. This disorder length scale (~ 1 nm) is much smaller than either the low-temperature electron mean free path or the Ginzburg-Landau superconducting coherence length of the system (both ~100 nm). Thus, the interdiffusion-induced disorder can be considered homogeneous on the relevant electronic lengths. Despite this, a highly inhomogeneous, patchy superfluid density with a typical length scale of micrometers has been observed in scanning probe measurements[25]. We associate this patchy superfluid density to the formation of superconducting puddles at the interface. The only other feature of the system with a similar large length scale is the domain structure in SrTiO$_3$ that forms below the cubic to tetragonal phase transition at 105 K. However, the spatial distribution of the superfluid density is clearly decoupled from the tetragonal domain boundaries, which follows high symmetry directions[25,26].

Therefore, we conclude that the formation of superconducting puddles here is a fundamental and emergent consequence of disorder. This is consistent with the results of computational studies indicating that microscopic disorder in 2D naturally gives rise to phase separation and thus emergent mesoscopic superconducting puddles[27,28]. Hence, we expect that inter-puddle phase coupling should play an essential role. In addition, we note that the features in the *R-T* data are remarkably similar to experiments of isolated superconducting islands on thin non-superconducting metallic films, where global phase coherence is mediated by inter-island (puddle) Josephson coupling[2,4,6,8]. Based on the combined analysis of magnetoresistance and current-voltage characteristics shown in the Supplementary Information, we interpret the first drop in resistance as the onset of pairing and formation of local emergent superconducting puddles ($T_P$: pairing temperature), consistent with the opening of a pseudogap[13]. With decreasing temperature, a second drop in resistivity occurs with the onset of long-range phase coupling limited by fluctuations ($T_F$: onset of macroscopic phase fluctuations). Further lowering temperature, the formation of global phase coherence leads to macroscopic 2D superconductivity ($T_C$: defined here as 1% $R_N$).

Extrapolating the characteristic temperatures (i.e. $T_P$, $T_F$, and $T_C$) to zero, we can now construct the full gate-dependent phase diagram as shown in Fig. 4**a**, representing gate-tuned ground states. From left to right in Fig. 4**a**, we see that the phase diagram crosses a rich set of distinct states: a "normal" state without pairing; a local superconducting puddle state without inter-puddle phase coupling; a phase fluctuating state in which macroscopic coherence is impaired by quantum fluctuations; and a global phase coherent 2D superconducting state. In particular, the system goes through a superconductor-metal transition when phase fluctuations destroy global coherence (across the dotted line). The dual electrostatic gates systematically control not only the interface carrier density but also the carrier distribution and the associated strength of interface disorder. $V_{TG}$ monotonically drives the proportional rise of $T_P$, $T_F$, and $T_C$ (Fig. 2**c**). This indicates that the increase of interfacial carrier density by $V_{TG}$ turns on superconducting pairing in local puddles and the interface progressively transitions into a macroscopic superconducting phase only after the puddles are fully phase coupled. On the other hand, the back-gate predominantly controls the sampling of interfacial disorder. This is highlighted by the diverging relationship between $T_P$ and $T_C$ on the negative side of the back-gate dome (Fig. 3**c**). With decreasing $V_{BG}$, the associated increase of the effective disorder induces stronger phase fluctuations among the puddles, leading to the collapse of $T_C$, although $T_P$ is increasing and remains finite. While the pairing strength ($T_P$) could be related to various factors including a gate tuned spin-orbit coupling[29], the 2D superconductor-metal transition below $T_P$ is governed by macroscopic phase fluctuations.

We emphasize here the distinction between the observed metallic state and a finite temperature crossover. In ordinary 2D metal thin films[30] that conform to the scaling theory of localization[15], the residual resistivity is much less than a quantum of resistance and has very weak temperature dependence. By contrast, the experiment here has a tunable resistivity at the lowest accessible temperature that can range from far lower to far above a quantum of resistance (for example in Fig. 2**a**), implying the presence of strong interactions. These interactions can create pronounced superconducting phase fluctuations, stabilizing metallic ground states over a wide range of parameters accessed here by dual electrostatic gating.

Finally, we note that many aspects of the phase diagram of Fig. 4 should be quite general for 2D superconductivity in the presence of disorder. The visibility and separation of the fluctuation regimes here are large due to several favourable characteristics. The superconducting gap is small (~ 40 μeV) such that disorder that is small relative to normal state scattering is still significant with respect to superconductivity. Furthermore, the carrier density is small, such that fluctuations on the scale of the superconducting coherence length are highly relevant. We note that very recent reports of gate-induced 2D superconductivity in boron nitride encapsulated van der Waals materials, such as monolayer $WTe_2$, are in a similar regime of low $T_C$ and low carrier density, and show a similar two-stage resistive transition for a range of gating into the superconducting state (Ref. 7, and Pablo Jarillo-Herrero, private communication). For different materials systems, the relative scale of disorder and density will determine the width and relevance of these intermediate fluctuation regimes, but they should be a generic feature of disordered 2D superconductivity.

**Acknowledgements** We thank S. A. Kivelson, A. Kapitulnik, Y. Suzuki, B. Spivak, P. A. Lee, K. Behnia, and N. Trivedi for discussions. This work was supported by the Department of Energy, Office of Basic Energy Sciences, Division of Materials Sciences and Engineering, under contract DE-AC02-76SF00515. Partial support was provided by the Stanford Graduate Fellowship in Science and Engineering (Z.C., H.I.), and the Gordon and Betty Moore Foundation's EPiQS Initiative through Grant GBMF4415 (low temperature measurements). A portion of this work was performed at the National High Magnetic Field Laboratory, which is supported by National Science Foundation Cooperative Agreement No. DMR-1157490 and the State of Florida. Part of this work was performed at the Stanford Nano Shared Facilities (SNSF).

**Author Contributions** Z.C. carried out the sample fabrication and experiments, and wrote the manuscript. A.G.S., H.Y., H.I., T.M., D.L., Y.X., H.Y., and Y.H. assist in measurements. H.Y.H. and Z.C. conceived the experiment. S.R. carried out theoretical analysis. All authors discussed the results and contributed to the manuscript.

**Author Information** The authors declare no competing financial interests. Correspondence and requests for materials should be addressed to zychen@stanford.edu or hyhwang@stanford.edu.

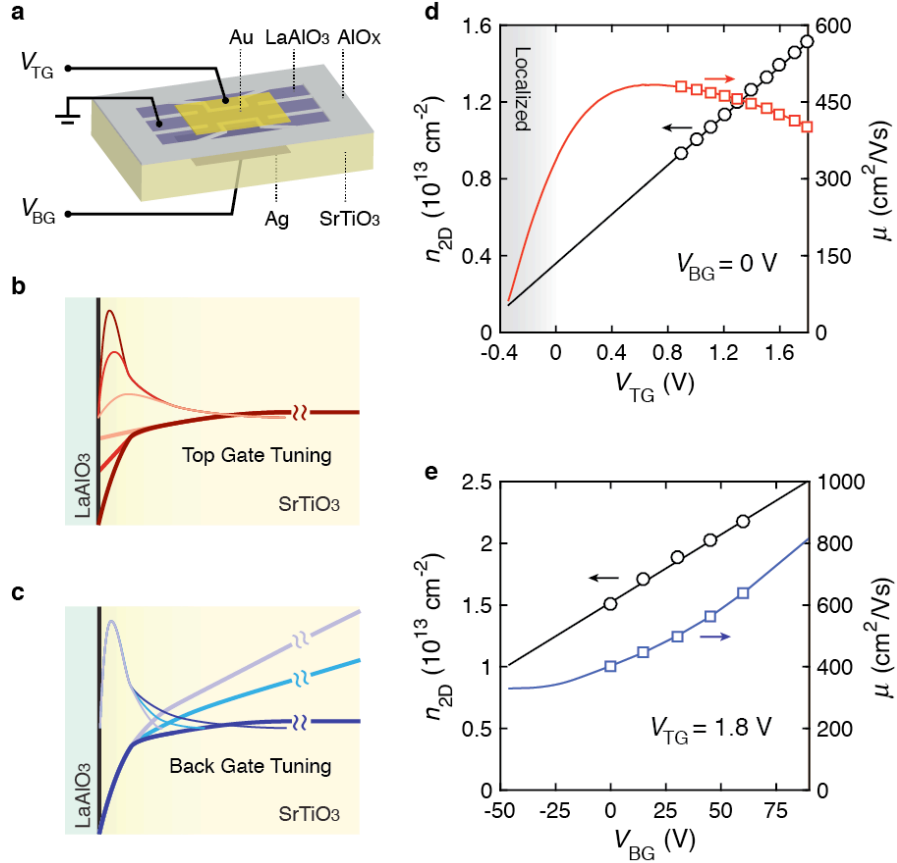

**Figure 1 | Normal state operation of the dual gate device. a**, The dual gate device is formed by applying a Au electrode on top of the 8 unit-cell epitaxial LaAlO$_3$ layer and a Ag electrode from the back of the SrTiO$_3$ substrate with 0.5 mm thickness. The channel width is 400 micrometers with Hall bars patterned by pre-deposited AlO$_x$ hard mask. **b,c**, Schematic diagram of the dual gate operation for top- and back- gates, respectively. The thin lines represent electron envelope wavefunction perpendicular to the interface at different gate voltages colour-matched to the thick solid lines, which represent the corresponding confinement potential. **d,e**, Top- and back- gate modulation of carrier density (circles) and Hall mobility (squares) at $T$ = 600 mK. The straight solid black lines are linear fits to the circles. The red and blue curves are estimates of mobility based on the measured resistivity and fitted extrapolation of Hall density. The ranges of the solid lines correspond to the experimentally accessible tuning ranges.

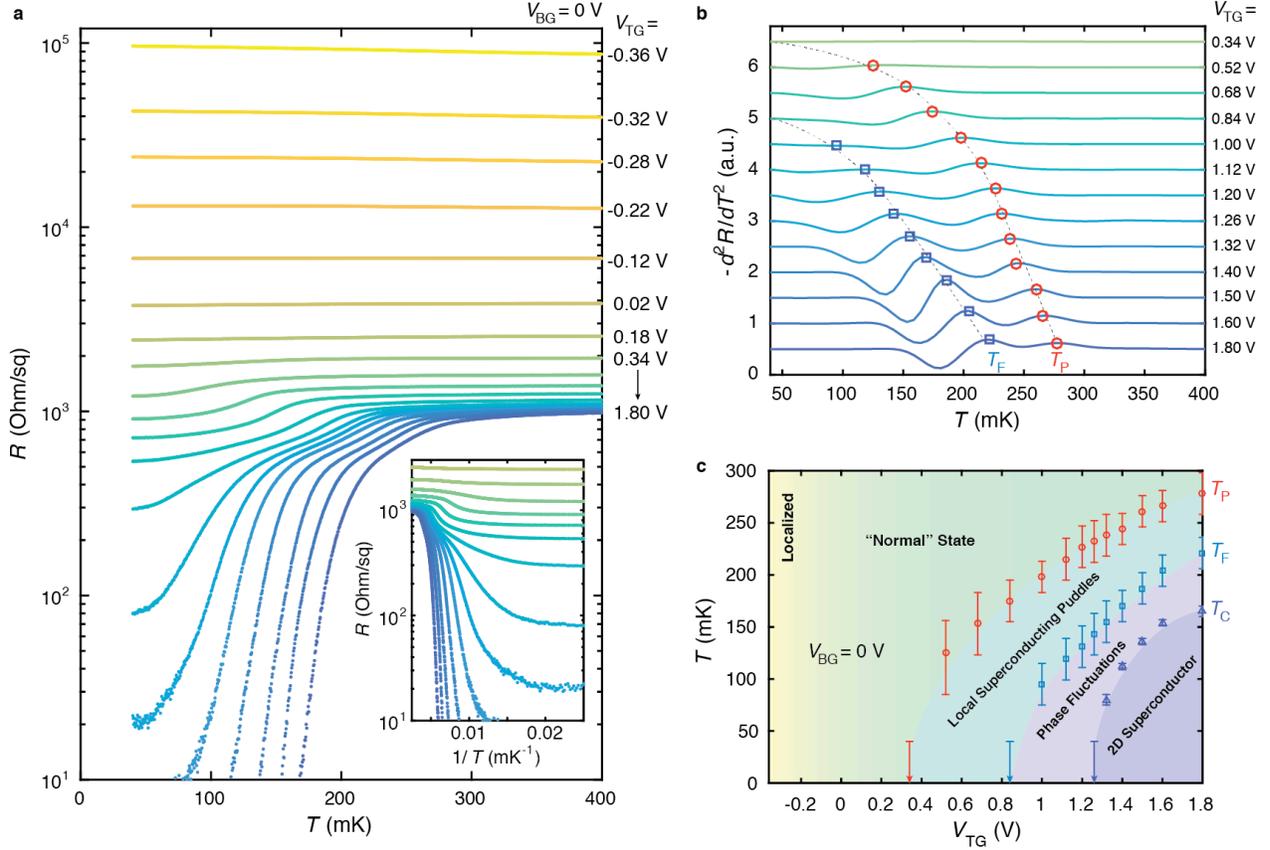

**Figure 2 | Top-gate modulation of the interface ground state. a**, Resistivity versus temperature (*R-T*) curves as a function of $V_{TG}$ from -0.36 V to 1.80 V with fixed $V_{BG}$ = 0 V. Inset: magnification of the data plotted against inverse temperature. **b**, The second order derivative calculated using a spline fit of the *R-T* curves shown in **a** for $V_{TG}$ from 0.34 V to 1.80 V. Red circles and blue squares indicate the local maxima of the peaks, defining $T_P$ and $T_F$. Curve colours are matched between **a** and **b**. Dashed lines are guides to the eye. **c**, Phase diagram extracted from **a** and **b**. Red circles, blue squares, and blue triangles represent $T_P$, $T_F$, and $T_C$, respectively. Error bars are defined by the full width of the second derivative peaks for $T_P$ and $T_F$. $T_C$ is defined as the temperature at which the resistivity drops to 1% of the normal state resistivity $R_N$ (400 mK). The lowest temperature measured for our *R-T* curves is 40 mK, so data points estimated to be lower than 40 mK are plotted with arrows.

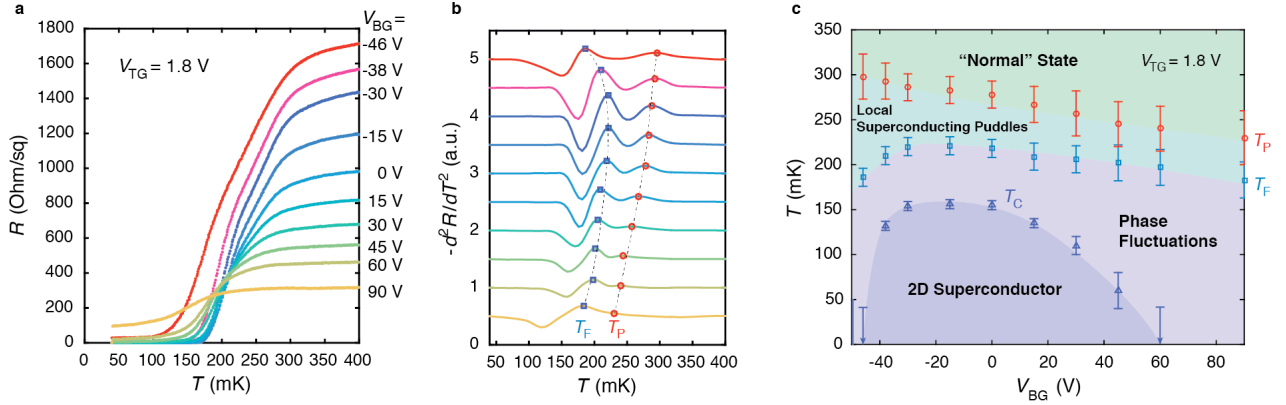

**Figure 3 | Back-gate modulation of the interface ground state. a**, Resistivity versus temperature (*R-T*) curves as a function of $V_{BG}$ from -46 V to 90 V with fixed $V_{TG}$ = 1.8 V. **b**, The second order derivative calculated using a spline fit of the *R-T* curves shown in **a**. Red circles and blue squares indicate the local maxima of the peaks, defining $T_P$ and $T_F$. Curve colours are matched between **a** and **b**. **c**, Phase diagram for back-gating with fixed $V_{TG}$ = 1.8 V. Red circles, blue squares, and blue triangles represent $T_P$, $T_F$, and $T_C$, respectively. Error bars are defined by the full width of the second derivative peaks for $T_P$ and $T_F$. $T_C$ is defined as the temperature at which the resistivity drops to 1% of the normal state $R_N$ (400 mK).

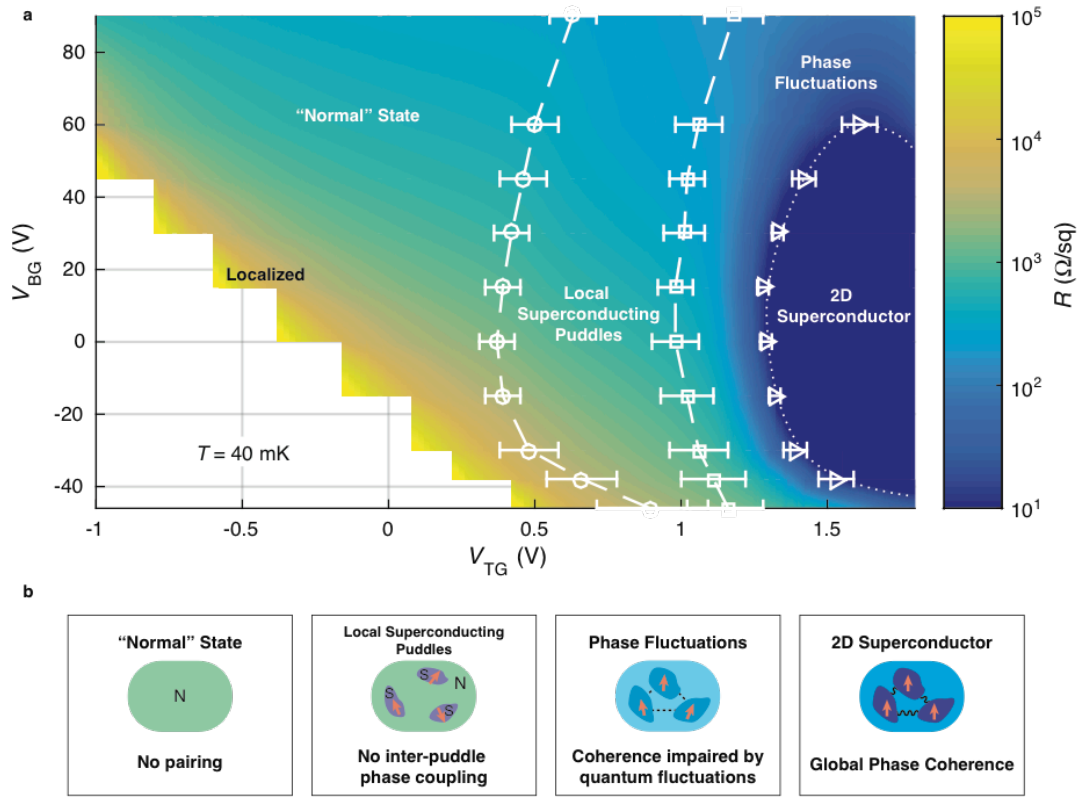

**Figure 4 | Ground state phase diagram by dual gates. a,** Resistivity mapping as a function of $V_{TG}$ and $V_{BG}$ measured at $T = 40$ mK. The colour scale is based on interpolated resistivity. Circles and squares are extracted from the peaks of the second order derivative calculated using a spline fit of the resistivity-versus-$V_{TG}$ curves. Error bars are defined by the full width of the second derivative peaks. Triangles are the gate voltages at which the resistivity drops to 1% of normal state resistance. The circles, squares, and triangles indicate the voltages where $T_P$, $T_F$, and $T_C$ extrapolate to zero temperature, respectively. The dotted line represents a superconductor-metal phase transition line. **b**, Schematics of the four different ground states in the phase diagram. The red arrows represent the superconducting phase of individual puddles. N: normal state. S: superconducting state.

## Methods

**Sample.** Ultraviolet photolithography was used to pattern an AlO$_x$ Hall bar hard mask (Fig. 1**a**; channel width 400 μm) on the SrTiO$_3$ substrate. The 8 unit-cell epitaxial LaAlO$_3$ was grown with pulsed laser deposition at 800 degree Celsius and $1 \times 10^{-5}$ Torr O$_2$, after pre-annealing at 950 degree Celsius in $5 \times 10^{-6}$ Torr O$_2$. After growth, the sample was post-annealed *in situ* at 600 degree Celsius for 3 hours at 0.2 bar O$_2$; this extended high-temperature post-annealing improved the quality of the LaAlO$_3$ epitaxial layer with respect to stability at high top-gate voltage. The gold top-gate electrode was deposited by electron-beam evaporation with a shadow mask. The silver back-gate electrode was applied by using silver conductive paste. Hall bar contacts were made with ultrasonic Al wire-bonding. In operation, DC voltages are applied to the top- and back- gates with respect to the channel.

**Magnetotransport measurements.** Resistivity measurements were performed using a 100 nA switched DC current. Strong nonlinear Hall effects are observed at $T = 600$ mK. To correctly estimate the total mobile carriers from the Hall effect, we used the high field limit of the slope (between 13 T and 14 T) for the Hall resistivity versus magnetic field curves as previously discussed[14].

## Data availability

The data that support the findings of this study are available from the corresponding author upon request.

# Supplementary Information

# The density and disorder tuned superconductor-metal transition in two dimensions

Zhuoyu Chen[1,2,3*], Adrian G. Swartz[1,2,3], Hyeok Yoon[1,2], Hisashi Inoue[1,2,3], Tyler Merz[1,2], Di Lu[1,3,4], Yanwu Xie[1,2,3], Hongtao Yuan[1,3], Yasuyuki Hikita[1,3], Srinivas Raghu[1,3,4], Harold Y. Hwang[1,2,3*]

[1]Geballe Laboratory for Advanced Materials, Stanford University, Stanford, CA 94305, USA.

[2]Department of Applied Physics, Stanford University, Stanford, CA 94305, USA.

[3]Stanford Institute for Materials and Energy Sciences, SLAC National Accelerator Laboratory, Menlo Park, CA 94025, USA.

[4]Department of Physics, Stanford University, Stanford, CA 94305, USA.

* zychen@stanford.edu; hyhwang@stanford.edu

## I. Introduction

In the supplementary information, we provide supporting evidence for the identification of $T_P$, $T_F$, and $T_C$ as defined in the resistivity-versus-temperature (*R-T*) curves in the main text. We measured magnetoresistance (MR) in perpendicular field and current-voltage (*I-V*) characteristics as functions of gate voltage and temperature, which are analyzed and presented in two sections as follows. In Section II, we show *I-V* characteristics as a function of $V_{TG}$ at $T = 40$ mK, in which the nature of the four regimes in the phase diagrams shown in the main text can be clearly distinguished. The system goes through a gate-tuned Berezinskii–Kosterlitz–Thouless (BKT) transition indicated by the power-law behavior in *I-V* curves. In Section III, we present a combined analysis of MR and *I-V* as a function of *T* and fixed gate voltages. The analysis of $H_{C2}$ extracted from MR yields the pairing temperature scale for superconductivity. The analysis of the power law exponent from *I-V* provides a BKT transition temperature, together with an onset temperature for superlinear power law. The correspondence between these extracted characteristic temperatures and the $T_P$, $T_F$, and $T_C$ as defined in *R-T* curves identifies these regimes from a different approach.

## II. Distinguishing different regimes by $V_{TG}$ tuned *I-V* characteristics

Figure S1 shows the *I-V* characteristics series corresponding to a horizontal cut in the $V_{TG}$ tuned phase diagram (Fig. S1**d**). Distinct *I-V* behavior in the four different regimes are shown in Figure S1**a**. The *I-V* curve for $V_{TG} = 0.2$ V exhibits linear behaviour in the entire current range, indicating a normal state response. For $V_{TG} = 0.6$ V, the *I-V* curve shows a partial drop of resistance at around 1.2 μA, but returns to linear dependence quickly below that, indicating finite superconducting pairing but only on a local scale. The *I-V* curve for $V_{TG} = 1.1$ V not only displays a partial drop of resistance at 2.2 μA, but also a power law behavior $V \propto I^\alpha$ below 1.0 μA, indicating the onset of long-range phase coupling governed by BKT physics. Yet further lowering the current, it returns to linear, indicating that the phase fluctuations coexist with dissipation. The *I-V* curve for $V_{TG} = 1.4$ V shows typical global 2D superconducting behaviour with voltage lower than the measurement noise limit for current lower than ~ 1.5 μA. The exponent *α* of the power law regime changes continuously as a function of $V_{TG}$ as shown in Fig.

S1**c**, crossing $\alpha = 3$ at $V_{TG} = 1.07$ V. This indicates that the system goes through a BKT-driven transition with fixed temperature $T = 40$ mK and varying $V_{TG}$.

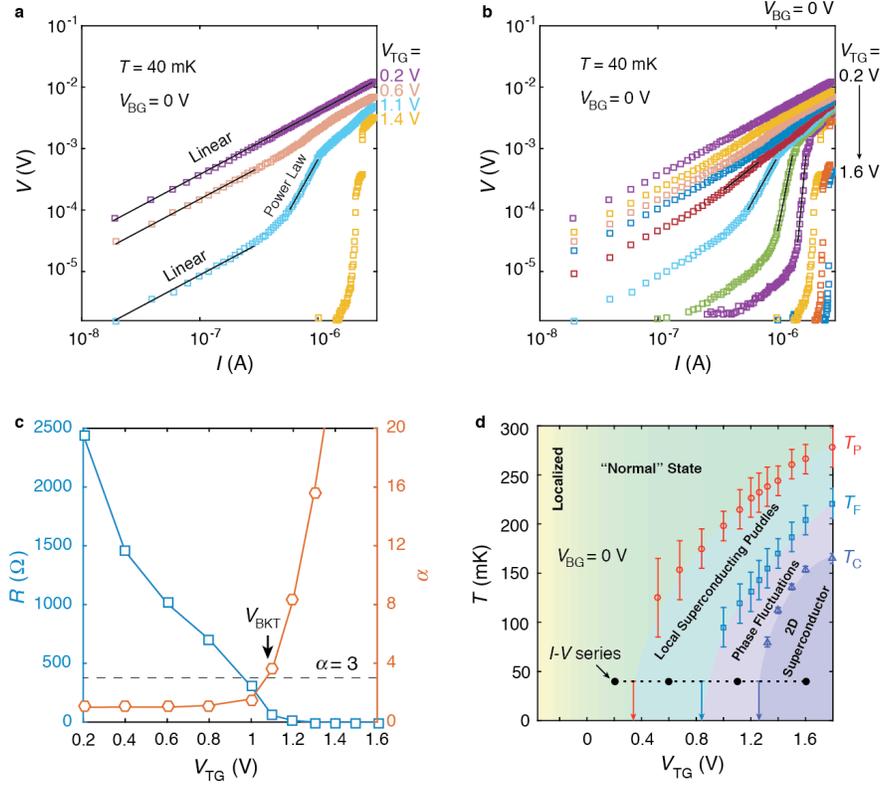

**Figure S1: a**, Four representative *I-V* curves in different regimes of the phase diagram as a function of $V_{TG}$ with fixed $V_{BG}$ (0 V) and $T$ (40 mK). **b**, *I-V* curve series with denser $V_{TG}$ intervals. Solid black lines indicate the power law regime in certain *I-V* curves. **c**, The left axis shows the resistance measured by the slope at the lowest currents as a function of $V_{TG}$. The right axis shows the exponent $\alpha$ of the power law regime as a function of $V_{TG}$. **d**, Reproduction of the phase diagram shown in Fig. 2**c** in the main text. The horizontal dashed line indicates where the *I-V* series is measured in the phase diagram, and the representative *I-V* curves in **a** are marked with solid circles.

### III. Nature of characteristic temperatures by combined analysis of MR and *I-V*

In this section, we show the correspondence between the characteristic temperatures defined in *R-T* curves (i.e. $T_P$, $T_F$, and $T_C$) and the characteristic temperatures extracted from MR and *I-V* analysis. This supports the identification of the different regimes in the phase diagram from *R-T* analysis.

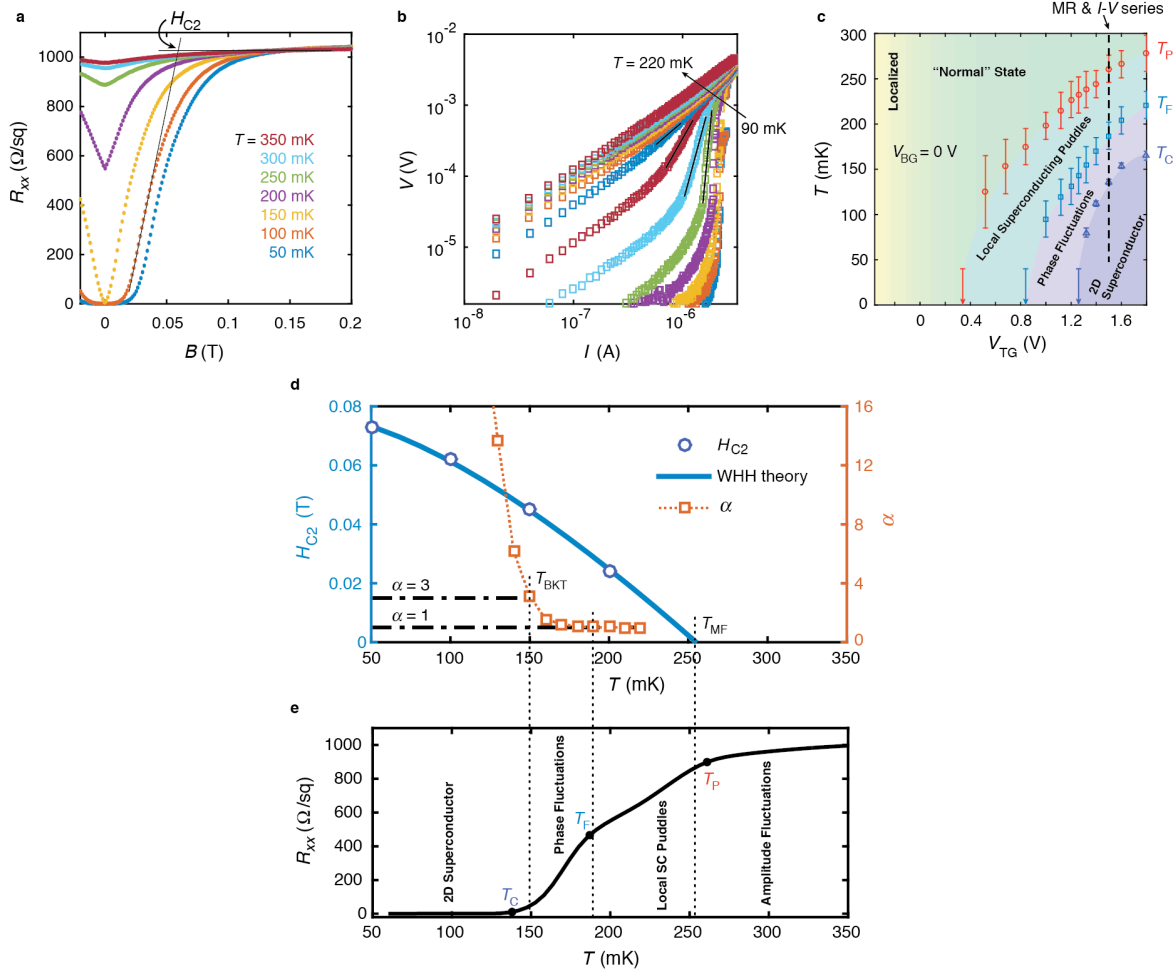

**Figure S2: a**, MR measured at different temperatures and fixed $V_{TG}$ =1.5 V, $V_{BG}$ = 0 V. The definition of $H_{C2}$ by crossing two fitted straight lines. **b**, *I-V* characteristics from $T$ = 90 mK to 220 mK with 10 mK intervals at fixed $V_{TG}$ =1.5 V, $V_{BG}$ = 0 V. The solid black lines fitted to the curves exhibit the regime for the power-law behavior. **c**, Reproduction of the phase diagram shown in Fig. 2c in the main text. The vertical dashed line indicates where the MR and *I-V* series are measured in the phase diagram. **d**, Combined results of the MR and *I-V* analysis. Blue circles represent $H_{C2}$ as defined in **a**. The solid blue curve is a fit with the Werthamer-Helfand-Hohenberg (WHH) form. Red squares represent power-law exponent $\alpha$ extracted from the *I-V* curves. Dashed horizontal lines indicate $\alpha$ = 3 and $\alpha$ = 1. **e**, *R-T* curve at $V_{TG}$ =1.5 V, $V_{BG}$ = 0 V. Solid circles show the characteristic temperatures $T_P$, $T_F$, and $T_C$ as defined in the *R-T* curve. Dotted lines exhibit the correspondence between these temperatures and the characteristic temperatures extracted from MR and *I-V* analysis as shown in **d**.

MR (Fig. S2**a**) in perpendicular field and *I-V* curves (Fig. S2**b**) are measured in a temperature series with fixed gate voltages, which corresponds to a vertical cut in the phase diagram shown in Fig. S2**c**. From the MR, we can experimentally define a characteristic field $H_{C2}$ as illustrated in the figure. The results shown in Fig. S2**d** suggest that the behaviour of the $H_{C2}$ can be fit by Werthamer-Helfand-Hohenberg (WHH) theory[1], which describes the mean-field behaviour of $H_{C2}$. Extrapolating the critical field $H_{C2}$ to zero, we can extract a mean field temperature $T_{MF}$ as a temperature scale for pair breaking. On the other hand, we can extract the exponent $\alpha$ from the power-law regime in the *I-V* curves. Shown in Fig. S2**d**, we can define the temperature at which $\alpha = 3$ to be the BKT transition temperature $T_{BKT}$. Interestingly, we found that the $T_{MF}$ closely matches $T_P$ defined in the *R-T* curve, as can be seen in Fig. S2**e**. $T_{BKT}$ is slightly higher than the $T_C$ extracted from the *R-T* curve (defined as 1% $R_N$), similar to the case for Fig. S1. Moreover, $T_F$ closely matches the temperature for the onset of a superlinear power law (i.e. from $\alpha = 1$ to $\alpha > 1$). These correspondences independently support the identification of $T_P$ as the pairing temperature; $T_F$ as the onset of macroscopic phase fluctuations; and $T_C$ for macroscopic phase coherence. The identification of the three characteristic temperatures divide the phase diagram into four regimes as discussed in the main text. Above $T_P$, the resistivity at zero field increases with increasing temperature, consistent with superconducting amplitude fluctuations. The MR in this regime can be well fitted with Aslamasov-Larkin (AL) theory[2] and Maki-Thompson (MT) theory[3,4], following similar procedures as in ref. 5.